# Multiferroic Core-Shell Nanofibers, Assembly in a Magnetic field and Studies on Magneto-Electric Interactions

G. Sreenivasulu,[1] Jitao Zhang,[1,2] Ru Zhang,[1] M. Popov,[1,3]  V. M. Petrov,[1,4] and G. Srinivasan[1]

[1] Physics Department, Oakland University, Rochester, MI 48309, USA

[2] College of Electrical and Information Engineering, Zhengzhou University of Light Industry, Zhengzhou 450002, People's Republic of China

[3] Faculty of Radiophysics, Electronics and Computer Systems, Taras Shevchenko National University of Kyiv, Kyiv, 01601, Ukraine.

[4] Institute of Electronic and Information Systems, Novgorod State University, Veliky Novgorod 173003, Russia

## ABSTRACT

Ferromagnetic-ferroelectric nanocomposites are of interest for realizing strong strain mediated coupling between electric and magnetic subsystems due to high surface area-to-volume ratio.  This report is on the synthesis of nickel ferrite (NFO) -barium titanate (BTO) core-shell nano-fibers, magnetic field assisted assembly into superstructures, and studies on magneto-electric (ME) interactions.  Electrospinning techniques were used to prepare coaxial fibers of 0.5-1.5 micron in diameter.  The core-shell structure of annealed fibers was confirmed by electron microscopy and scanning probe microscopy.  The fibers were assembled into discs and films in a uniform magnetic field or a field gradient.  Studies on ME coupling in the assembled films and discs were done by magnetic field $H$ induced polarization, magneto-dielectric effects at low frequencies and at 16-24 GHz, and low frequency ME voltage coefficients (MEVC).  We measured

~ 2-7% change in remnant polarization and in the permittivity for $H = 7$ kOe, and a MEVC of 0.4 mV/cm Oe at 30 Hz.  A model has been developed for low-frequency ME effects in an assembly of fibers and takes into account dipole-dipole interactions between the fibers and fiber discontinuity. Theoretical estimates for the low-frequency MEVC have been compared with the data. These results indicate strong ME coupling in superstructures of the core-shell fibers.

**1. Introduction**

Materials with simultaneous long range ordering of magnetic and electric dipoles of are multiferroic. Single-phase multiferroics that show evidence for coupling between the magnetic and electric subsystems were investigated extensively in the past [1-5]. Measurements of the magneto-electric (ME) coupling strength generally involve magnetic field $H$ induced polarization $P$, magneto-dielectric effect ($H$ induced variation in the dielectric constant), and electric field $E$ induced magnetization, anisotropy field or permeability. A majority of the single-phase multiferroics show weak ME coupling even at very low temperatures [1]. In recent years, however, several single-phase materials including $BiFeO_3$ and M- and Z-type hexagonal ferrites were reported to show strong ME coupling [5-7].

Strong ME interactions at room temperature could be realized in multiferroic composites [8-12]. A ferromagnetic-ferroelectric composite, for example, is expected to show coupling between the ferroic phases that is aided by mechanical strain. The ME coupling in this case arises due to magnetostriction in the ferromagnetic phase and piezoelectric effect in the ferroelectric phase, leading to an electrical response in an applied magnetic field $H$ or a change in magnetization or anisotropy field in an electric field $E$. Such composites made by cosintering ferrites and barium titanate were first studied in the 1970s and the ME coupling was found to be rather weak due to leakage currents [13, 14]. Layered composites with alternate layers of low-resistivity ferrites and high resistivity ferroelectrics were investigated in the 1990s and were found to show strong ME coupling [15,16]. Several layered composites with ferromagnetic metals/alloys and ferroelectric BTO or lead zirconate titanate (PZT) also showed strong ME interactions [8].

Very recent efforts on enhancing the ME coupling in multiferroic composites have focused on nanocomposites [17-27]. Since the ME coupling originates from strain transfer at the interface between the ferroic phases, nano-composites with surface area-to-volume ratios that are orders of magnitude higher than bulk or layered systems are expected to show very strong ME coupling. Synthesis of ferrite-ferroelectric core-shell nanoparticles by chemical- or DNA-assisted self-assembly and measurements of ME coupling were recently reported [23-27]. Evidence for strong ME coupling, for example, was obtained in nickel ferrite-BTO for core-shell particles. Past efforts on multiferroic nanofibers include synthesis of ordered ferrite-piezoelectric core-shells nanotubes on porous anodized aluminum oxide or membrane templates by a combined sol-gel and electrochemical deposition technique or by a two-step sol-gel process [28-30]. Free-standing core-shell tubes were obtained by dissolving the templates (in NaOH for example). Fibers with nanocrystals of ferrite and PZT were made by gel-thermal decomposition and electrospinning was used for fibers with alternate layers of ferrite and PZT or coaxial fibers [31-33]. We recently reported on the synthesis of nickel ferrite-PZT core-shell nanofibers by electrospinning and measurements of ME coupling by $H$-induced polarization and magneto-dielectric effects [34].

This work in on the synthesis of nanowires of nickel ferrite, $NiFe_2O_4$ (NFO) and barium titanate $BaTiO_3$ (BTO) by electrospinning, assembly into films and discs in a magnetic field and measurements on ME interactions. The coaxial fibers were characterized by scanning electron microscopy, X-ray diffraction, and scanning probe microscopy. Fibers free of impurities with uniform core and shell structures were evident from these studies. Ferromagnetic and ferroelectric order parameters for the fibers compare favorably with parameters for bulk NFO and BTO. A strong ME coupling between the electric and magnetic subsystems was inferred from measurements of magnetic field induced polarization and magneto-dielectric effect in discs of fibers pressed in a uniform magnetic field. Low-frequency ME voltage coefficient (MEVC) measurements were carried out on films assembled in a magnetic field gradient. A maximum MEVC of ~ 0.4 mV/cm Oe was measured at 30 Hz. A model was developed for the low-frequency

ME coupling in the assembled films. We considered dipole-dipole interactions and porosity due to fiber discontinuity in the model and estimated ME coupling compare favorably with the data. Details on the synthesis, structural and ferroic order parameter characterization, assembly of fibers in magnetic fields, results on ME coupling and our theory are discussed next.

## 2. Experiment

The procedure for the synthesis of coaxial fibers of nickel ferrite ($NiFe_2O_4$) and barium titanate ($BaTiO_3$) involved the preparation of sol-gel for the two oxides and fiber synthesis by electrospinning with the use of a dual syringe pump and a coaxial needle [34-37]. The preparation of the sol-gel was carried out individually for nickel ferrite and BTO. The NFO sol was prepared with 1.2g of poly(vinylpyrroli-done) (PVP, MW~1,300,000) and was dissolved in a mixture of ethanol (13ml) and distilled water (7ml) followed by magnetic stirring for 1h to ensure the dissolution of PVP. Then 1.767g of nickel (II) acetate tetrahydrate [$Ni(CH_3COO)_2.4H_2O$], and iron (III) nitrate nanohydrate [$(Fe(NO_3)_3.9H_2O)$] with 1:2 molar ratios of Ni:Fe were added to the PVP/ethanol/water solution and further magnetically stirred for about 20 hrs at room temperature to form a homogeneous viscous solution with PVP concentration of 6 wt.% for electrospinning [34,37]. The procedure for the preparation of sol-gel of $BaTiO_3$ is as follows [36,37]. Barium acetate and titanium isopropoxide were used. First, 1.275 g of $Ba(CH_3COO)_2$ was dissolved in acetic acid and stirred. Then 1.475 ml of [$(CH_3)_2CHO]_4Ti$ was added under continuous stirring. The solution was then mixed with a solution of PVP dissolved in ethanol. After stirring for several hours at room temperature, the solution was used for electrospinning.

Our electrospinning system consisted of a dual syringe pump (NE-4000, New Era Pump Systems, Inc.) and a coaxial stainless steel needle (rame-hart) with inner and outer diameters of 400 μm and 1050 μm, respectively, and wall thickness of 150 μm. A high voltage power supply (PS375/+20kV, Stanford Research Systems, Inc.) was used to apply the necessary voltage to the needle. The grounded collector was an aluminum drum placed 8-15 cm from the tip of the needle. Both BTO and NFO solutions were placed in 10 mL syringes, attached to the syringe pump and the sol was fed into the coaxial needle at a flow rate of 0.3 mL/h. The aluminum drum was rotated at an optimum speed to collect the fibers under a DC voltage of 15-20 kV. Humidity in the electrospinning chamber was maintained in the range 30-40%. The fibers were collected, dried in an oven at 40 C for 24 hrs and then annealed for 1 h at 600 - 700°C in air. The heating and cooling rates for annealing was 1°C/min.

Two types of fibers were made, BTO core-NFO shell (Sample-A) and NFO core-BTO shell (Sample-B). The structural and chemical composition characterization of the fibers were carried out using an X-ray diffractometer (XRD) with Cu Kα radiation, a scanning electron microscope (JEOL JSM 6510 SEM) and a scanning probe microscope (Park Systems XE-100E). Energy dispersive x-ray spectroscopy (EDS) was used for chemical composition determination. Ferroelectric characterization was carried out with a Radiant Ferroelectric Tester. Ferromagnetic resonance measurements at 2-20 GHz were done with a microstripline transducer on an alumina ground plane and a vector network analyzer. Magneto-electric characterization was performed by (i) static magnetic field induced polarization, (ii) magneto-dielectric effect at 16-24 GHz, and (iii) low-frequency ME voltage coefficients.

### 3. Results and Discussion

*3.1. Structural characterization*

The fibers annealed at high temperatures were imaged by scanning electron microscopy (SEM)

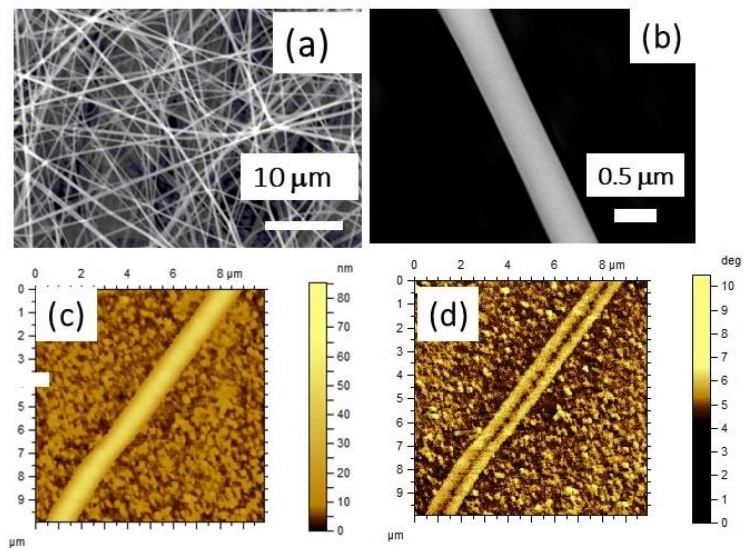

Fig.1. SEM micrograph of (a) coaxial fibers of Sample A with BTO core and NFO shell and (b) a single fiber of Sample A. (c) AFM topography for fibers of Sample B with NFO core-BTO shell. (d) MFM phase image for Sample-B.

and scanning probe microscopy (SPM) techniques and are shown in Fig.1 for Samples A and B. The SEM micrograph in Fig.1(a) for Sample-A shows well formed fibers with a distribution in the diameter varying in the range 300-700 nm with 60% of the fibers with a diameter of 500±50 nm. Figure 1(b) shows SEM image of a single fiber of Sample-A with a diameter of 500 nm. Sample-B with ferrite core and BTO shell had a much larger diameter compared to Sample-A and ranged from 1200 nm to 1800 nm with more than 50% of the fibers with a dimeter of 1500±100 nm. A powerful tool for the fiber characterization is the magnetic force microscopy that allows imaging of the magnetic response from NFO. AFM topography and MFM phases images for Sample-B are shown in Fig.1 (c) and (d). The MFM phase image clearly shows the NFO core and BTO shell in different contrasts. The core and the shell are well resolved in the MFM image. The SPM measurements on Sample-A showed fibers with an average shell diameter of 500 nm and core diameter of 250 nm and lengths 10-100 μm, corresponding to volume fractions of 25% and 75% for BTO and NFO, respectively. Sample-B had a much higher shell diameter of 1500 nm and a core diameter 500 nm. The corresponding volume fraction for NFO and BTO are 12% and 88%, respectively. Any difference in viscosity of sols of NFO and BTO could account for the large difference in the diameters for the two types of fibers.

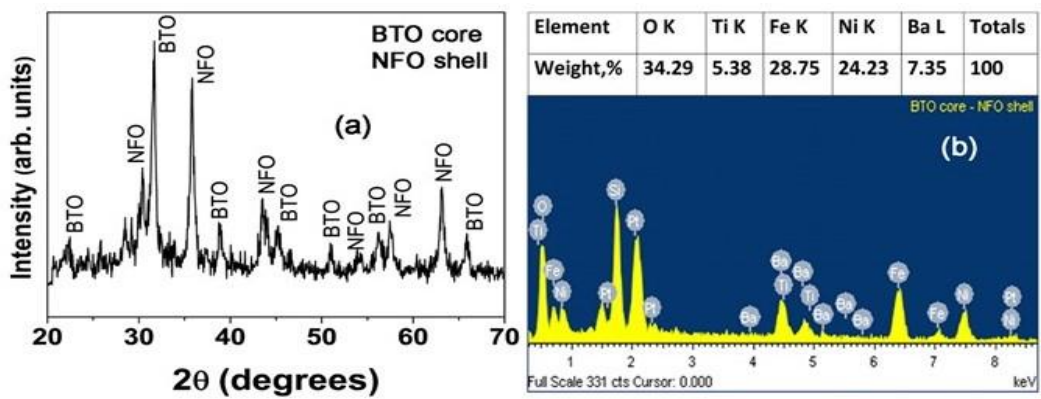

Fig.2. X-ray diffraction and energy dispersive x-ray diffraction data for Sample A.

Figure 2 shows results of X-ray diffraction and chemical composition analysis by energy dispersive X-ray spectroscopy (EDS). The fibers showed XRD peaks corresponding to BTO and NFO and were free of any impurity phases. The chemical composition for the fibers estimated from EDS data was in good agreement with the expected composition.

### 3.2. Ferroic order parameters

The fibers were characterized in terms of ferromagnetic and ferroelectric order parameters. The ferroelectric nature of the fiber was investigated by polarization $P$ vs. $E$ measurements. A (Radiant) ferroelectric tester was used. The fibers were mixed with a small amount of binder and pressed into a disk for these measurements. Figure 3 shows the $P$ vs $E$ data for both samples. One observes the expected ferroelectric behavior with remnant polarization $P_r = 1$ $\mu C/m^2$ and 2 $\mu C/m^2$ for samples A and B, respectively. Sample B with BTO volume fraction of 88% shows a larger $P$-value compared to Sample A with 25% BTO. But the $P$-values for the fibers are smaller than reported values for bulk or thin film BTO [35,36].

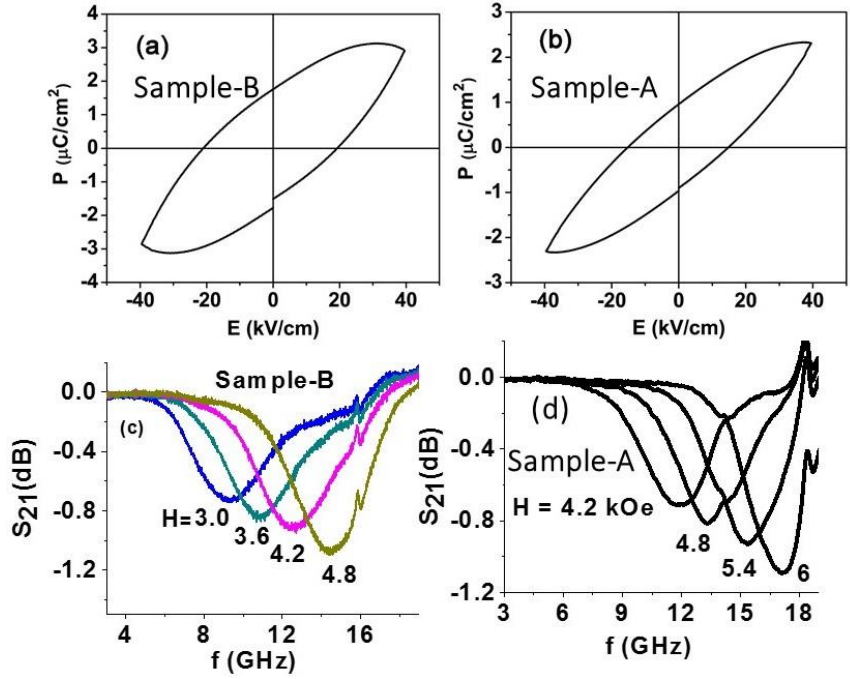

Fig.3. (a) *P* vs *E* for disc of fibers with NFO core-BTO shell (Sample B). (b) Similar data as in (a) for fibers with BTO core-NFO shell (Sample A). (c) Scattering matrix parameter $S_{21}$ versus frequency *f* profiles showing FMR for a series of bias fields *H* for a pressed rectangular disc of Sample B. (d) Similar FMR profiles for Sample-A.

Ferromagnetic resonance (FMR) technique was utilized for the magnetic characterization of the annealed fibers. A rectangular sample of dimensions 3 mm × 3 mm × 0.2 mm made by pressing annealed fibers was mounted in a test structure with a stripline transducer on an alumina ground plane and excited with microwave power. A static magnetic field *H* was applied parallel to the sample plane. Figure 3 shows the scattering parameter $S_{21}$ vs frequency *f* profiles for a series of *H*-values. FMR manifests as a dip in the transmitted power as seen in the profiles. Data on resonance frequency $f_r$ as a function of *H* were obtained from the profiles and was fitted to the resonance condition $f_r^2 = \gamma\, H\, (H + 4\pi M_{eff})$ where $\gamma$ is the gyromagnetic ratio and $4\pi M_{eff}$, the effective magnetization, is given by $4\pi M_{eff} = 4\pi M_s + H_a$. Here $4\pi M_s$ is the saturation induction and $H_a$ is the anisotropy field. From the data in Fig.3(c) the estimated values of magnetic parameters for Sample-B are $\gamma = 3.1$ GHz/kOe and $4\pi M_{eff} = 400$ G. The $\gamma$-value is in agreement with reported value for nickel ferrite [38]. Since the ferrite volume fraction in Sample B is 12 %, and assuming negligible anisotropy field $H_a$, the corresponding ferrite-only $4\pi M_s$ ~ 3.3 kG which is also in agreement with reported values for NFO [38]. Similar FMR results were obtained for sample A and are shown in Fig.3 (d). The estimated magnetic parameters from the FMR data are $\gamma = 3.1$ GHz/kOe and $4\pi M_{eff} = 480$ G. Since the ferrite volume fraction in Sample-A is 75% the effective magnetization corresponds to $4\pi M_s$ of 640 G that is quite small compared to magnetization for bulk NFO. One may, therefore, infer that anisotropy field is quite high in Sample-A compared to Sample-B. The possible origin of the large anisotropy field could be the magnetic dipole-dipole interaction between fibers with ferrite shells in Sample-A. Such dipole-dipole interactions are expected to be rather weak in 1500 nm diameter fibers of Sample-B with ferrite core and BTO shell.

*3.3 Magnetoelectric interactions*

*(i) H-induced polarization*: The strength of direct magneto-electric (DME) effects in the annealed, magnetic field oriented fibers were investigated by measurements of ferroelectric hysteresis under a static magnetic field. Samples A and B for the measurements were made by adding a binder to the fibers, loaded on to an alumina die and exposed to a static magnetic field of 7 kOe. The oriented fibers were then pressed into a circular disc in the dye with a hydraulic press. The disc was placed in the ferroelectric tester. A static magnetic field $H$ was applied parallel to the sample plane. Data on $P$ vs. $E$ as in Fig.3 were obtained for $H = 0$ to 7 kOe for both increasing and decreasing $H$-values and the change in the remnant polarization $P_r$ under $H$ was determined. Figure 4 shows the fractional change in $P_r$: $\Delta P_r / P_r (H=0) = [P_r(H)-P(H=0)]/P_r (H=0)$ for $H = 0$-7 kOe. Data in Fig.4(a) for Sample A with BTO core and NFO shell shows an initial increase in $\Delta P_r/P_r$ with $H$ up to 5 kOe and then it decreases rapidly to zero for $H = 7$ kOe. Upon decreasing $H$ from 7 kOe to zero, $\Delta P_r/P_r$ becomes negative with a maximum value of 1.5% for $H$=0. A somewhat different character is seen in Fig.4(b) for Sample-B with NFO core and BTO shell. An increase in $\Delta P_r/P_r$ is observed as $H$ is increased from 0 to 7 kOe. With subsequent decrease in $H$ back to zero, $\Delta P_r/P_r$ shows further increase to a maximum value of ~7% for $H$=0. Thus it is evident from the results in Fig.4 that a strong strain mediated ME coupling is present in the core-shell fibers and that the nature of coupling depends on the ferroic phases in the core and the shell. The data in Fig.4 compare favourably with $H$-induced change in polarization reported for core-shell fibers of NFO and PZT [34].

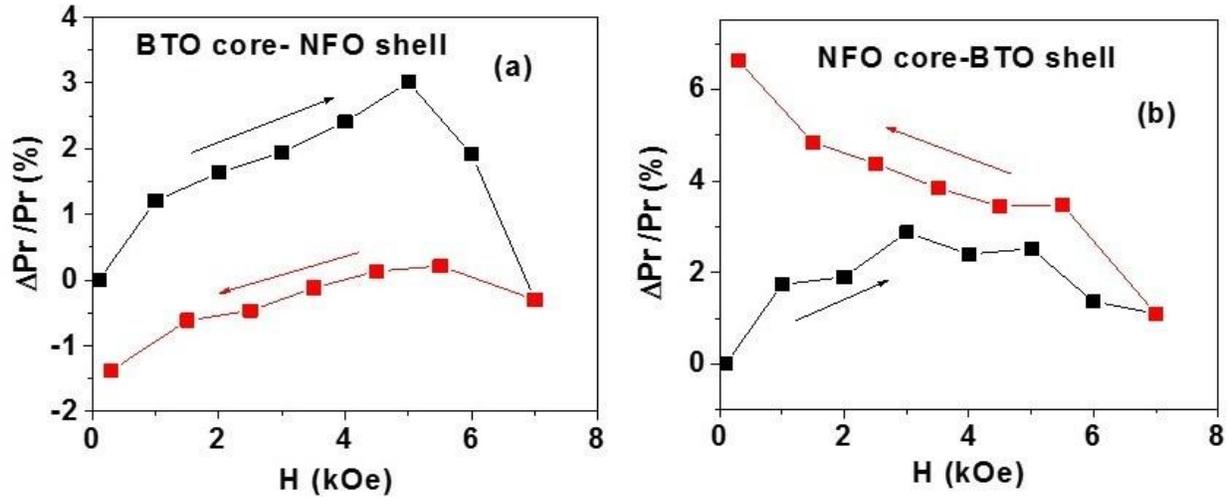

Fig.4. Fractional change in the remnant polarization as a function of static magnetic field $H$ for discs of magnetic field oriented annealed fibers of (a) Sample-A and (b) Sample-B.

*(ii) Magneto-dielectric effect*: We carried out measurements of $H$- induced variations in dielectric permittivity $\varepsilon$ versus frequency $f$ on discs of fibers oriented in a magnetic field. Measurements were done at low frequency, 50-200 Hz, with an RLC meter and at 16-24 GHz using a vector network analyser. Figure 5 shows the real part of the permittivity $\varepsilon'$ vs $f$ for Sample-B. For $H$=0 a decrease in $\varepsilon'$ with increasing $f$ is measured. With the application of $H$=3 kOe, $\varepsilon'$ decreases and the fraction decrease in $\varepsilon'$ shown in Fig.5(b) ranges from 1 % to 2.3%.

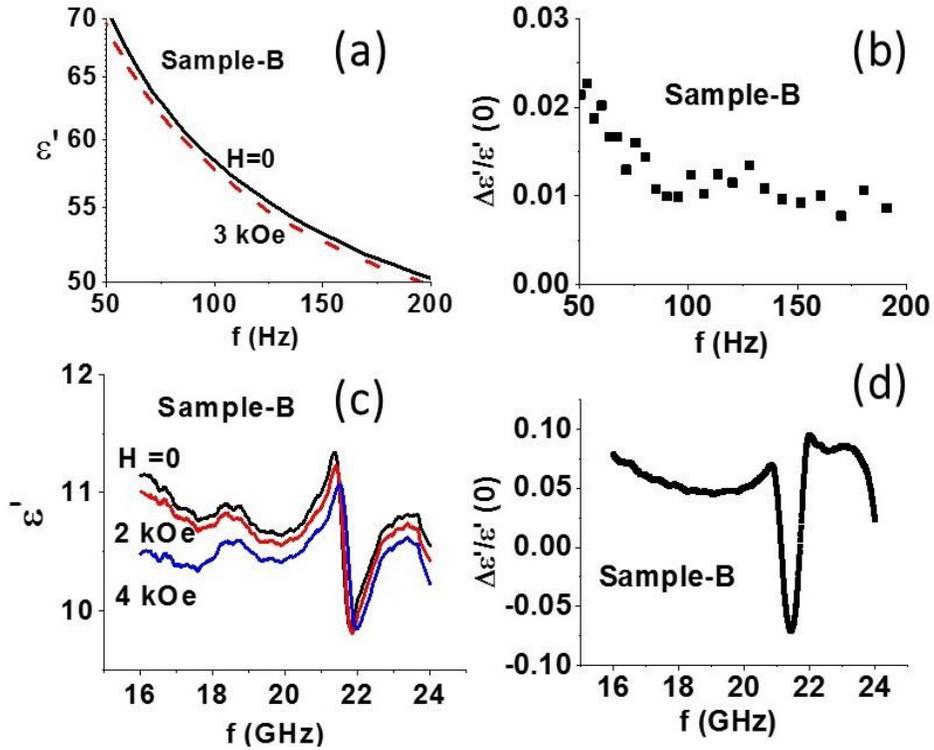

Fig.5. (a) Frequency dependence of the real part of the relative permittivity $\varepsilon'$ for discs of NFO-BTO fibers. Data are for static magnetic field $H$ parallel to the disc plane. (b) Estimated fractional decrease in $\varepsilon'$ vs. $f$ from data in Fig.5(a). (c) Data on $\varepsilon'$ vs. $f$ for a series of $H$ for a rectangular disc sample of NFO-BTO core-shell fibers oriented in a magnetic field. The sample dimension was chosen so that dielectric resonance is present in the sample in the above frequency range. (d) ) Estimated fractional change in $\varepsilon'$ vs. $f$ from data in Fig.5(c).

Similar studies on magneto-dielectric effects were done for the frequency range 16 to 24 GHz on rectangular discs of magnetic field oriented fibers. The transmission line method used involved placing the sample inside a rectangular waveguide. A precision quarter-wavelength WR-42 waveguide section was used as the sample holder. A pellet of the fibers with lateral dimensions 10.7 mm × 4.3 mm that filled the cross section of the waveguide without gaps was used. The sample thickness was in the range 0.4 to 0.6 mm. The waveguide with the sample was excited with microwave power and a two-port measurement of transmission and reflection coefficients were done to estimate the complex permittivity. During the measurements a bias static magnetic field was applied along the wide wall of waveguide, and dielectric constant variation with frequency and H was recorded for H=0-4 kOe.

Data on $\varepsilon'$ vs $f$ for a series of $H$ over the range 16-24 GHz are shown in Fig. 5(c). A resonance is clearly seen in the data and is due to dielectric resonance in the sample. With the application of $H$, a general decrease in $\varepsilon'$ is observed except for a narrow frequency range extending from 20 to 21 GHz over which $\varepsilon'$ is found to increase with increasing $H$. The fractional change in the dielectric constant for $H$=4 kOe is shown in Fig.5(d). For frequencies away from the resonance region one measures a decrease $\varepsilon'$ and a fractional change of 5 to 7% in $\varepsilon'$. For the frequency range 20-21 GHz close to resonance the dielectric constant increases in an applied field and the fractional change is -5 to -7 %. Thus a giant magneto-dielectric effect in the core-shell fiber composite with a 14% net change in permittivity at around the dielectric resonance is evident in the data of Fig.5(d).

*(iii) Low-frequency ME voltage coefficient*: Next we discuss magnetic field assisted assembly (MFAA) of the fibers into films for ME voltage coefficient measurements. Two types of assembly were done: (i) in a uniform field produced by a solenoid or an electromagnet that is expected to align them along the field direction and (ii) a non-uniform field produced by a permanent magnet which will exert an attractive force on the fibers and move them toward regions of high field strength [39-41]. Fibers were mixed with a solvent and dispersed on a glass slide. A solution with 25 mg of the fibers, 5 mg of PVP and 1ml of methanol was ultrasonicated for uniform mixing and then dispersed on a 10 mm × 5 mm × 0.5 mm glass plate. The glass slide was placed between the pole pieces of an electromagnet and subjected to fields of 3-5 kOe for assembly in a uniform field. Figure 6 shows the SEM micrograph of the resulting superstructure for fibers of Sample-A. Fibers are aligned parallel to the field direction. When the fiber concentration is high enough they form a planar structure consisting of dense linear chains as in Fig.6(a).

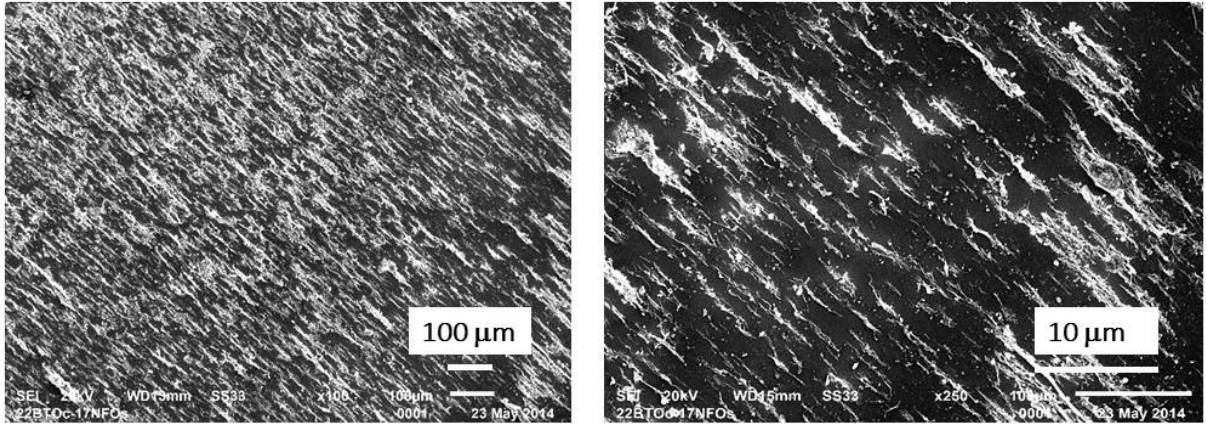

Fig.6. SEM micrograph for core-shell fibers of Sample-A assembled in a uniform magnetic field (left) and in a non-uniform field produced by a permanent magnet (right).

For assembly in a non-uniform field, the glass slide was placed on a (barium ferrite) permanent magnet that produced a field gradient of 400 Oe/cm. The force exerted by the field gradient resulted in regions of high concentration of parallel fibers as shown in Fig.6(b).

The magnetic-field-assembled films were characterized for the strength of the direct-ME effects. The low frequency ME voltage coefficient (MEVC) measurements were done on films assembled in a uniform magnetic field. The films were assembled between two parallel electrodes of 30 nm Ti-3 μm Pt separated by 4 mm on a glass slide of lateral dimensions 10 mm x 8 mm. Fibers in a solvent such as ethanol were dispersed between the electrodes, subjected to an in-plane field of 5 kOe so that a film of 0.2 mm in thickness was formed as the solvent evaporated. For ME measurements the films were subjected to a DC bias magnetic field $H$ and an ac field of $H_{ac} = 1$ Oe at 30 Hz, both parallel to the electrodes (perpendicular to the array of parallel chains as in Fig.6a). The ME voltage induced across the electrode was measured as a function of $H$. Figure 7 shows MEVC vs $H$ data for films of Samples A- and B. Consider first the results in Fig.7(a) for the film with fibers of Sample-A. The MEVC shows a zero-bias ($H$=0) value of

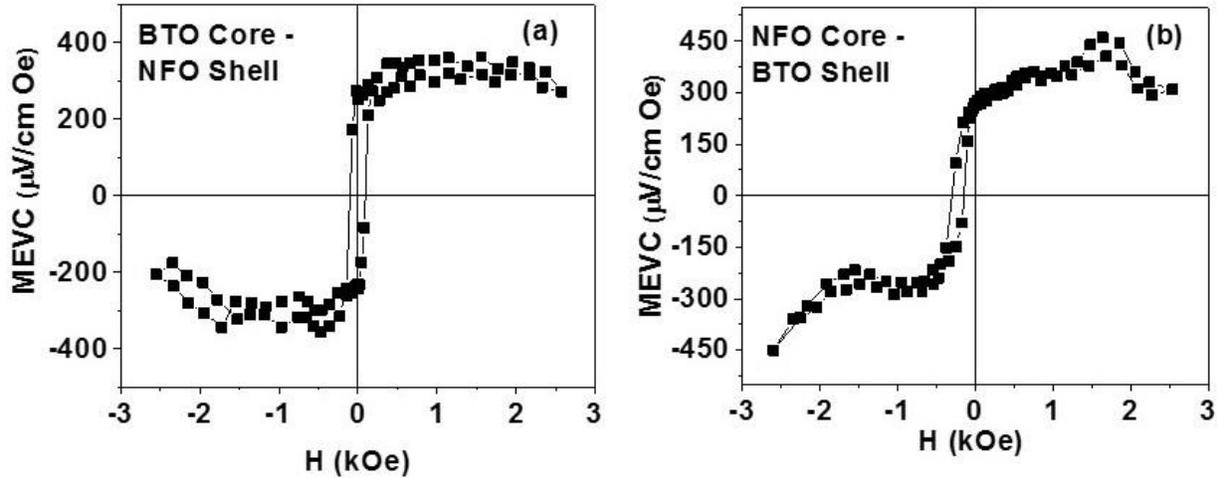

Fig.7. Low-frequency ME voltage coefficient measured in magnetic field assembled films of (a) Sample-A and (b) Sample-B.

300 μV/cm Oe and is indicative of a built-in magnetic field possibly arising from dipole-dipole interactions between NFO shells in the core-shell structures. With an increase in $H$, MEVC increases to a maximum value of 350 μV/cm Oe. The coupling strength remains the same for $H$ up to 2.5 kOe. A hysteresis is also seen in MEVC vs $H$ data. When $H$ is reversed the MEVC becomes negative (a $180^0$ phase difference with $H_{ac}$). Data in Fig.7(b) for the films with fibers of NFO core and BTO shell show similar features as for Sample-A. But the zero-bias MEVC is slightly smaller than for Sample-A. The maximum MEVC is 450 μV/cm Oe. The overall strength of low-frequency ME effects is the same for both samples although the volume fraction of ferrite and ferroelectric phases in the two samples are quite different.

*(iv) Model for low-frequency ME effects*: Theories for strain mediated ME coupling in bulk, layered and nanostructures of ferromagnetic-ferroelectric composites were proposed in the past [15,42,43]. The data in Fig.7 for magnetic field assembled films of core-shell fibers, however, require modifications to existing models. It is necessary to take into account electric and magnetic dipole-dipole interactions between fibers and fiber discontinuity in an assembled structure. We consider a fiber in the (*1,2*) plane with the length *L* (along direction *3*) greater than its radius. In this case, demagnetization effects vanish for bias magnetic field and ac field along the fiber axis. The theory of ME coupling in a core-shell structure is similar to that in a free-standing nanopillar [43]. Using the elastic and constitutive equations for a BTO core of radius $^pR$ and NFO shell of radius $^mR$, and the boundary conditions enables one to obtain the *1-D* approximation of ME voltage coefficient in explicit form:

$$\alpha_{E33} = \frac{^pd_{33}{}^mq_{33}r_p^2(1-r_p^2)}{[^p\varepsilon_{33}r_p^2 + {}^m\varepsilon_{33}(1-r_p^2)][^ps_{33}(1-r_p^2) + {}^ms_{33}r_p^2]} \qquad (1)$$

where $q_{33}$ and $d_{33}$ are piezomagnetic and piezoelectric coupling coefficients, $\varepsilon_{33}$ is the permittivity, $s_{33}$ is the compliance coefficient, and $r_p = {}^pR/{}^mR$. The superscripts *p* and *m* refer to piezoelectric and magnetic phases, respectively. For a fiber with the magnetostrictive core and piezoelectric shell, expression for ME voltage coefficient is given by:

$$\alpha_{E33} = \frac{^pd_{33}{}^mq_{33}r_m^2(1-r_m^2)}{[^p\varepsilon_{33}(1-r_m^2) + {}^m\varepsilon_{33}r_m^2)][^ps_{33}r_m^2 + {}^ms_{33}(1-r_m^2)]} \qquad (2)$$

where $r_m = {}^mR/{}^pR$.

Estimates of low- frequency ME voltage coefficient as a function of the ratio of core-to-shell radius for the fibers of Sample-A are shown in Fig.8(a). For fibers with BTO core and NFO-shell estimated MEVC in Fig. 8(a) shows a maximum value of 1.6 V/cm Oe for $r_p = 0.33$. Similar estimates of MEVC vs $r_m$ for Sample-B are shown in Fig.8(b) and the maximum ME response is predicted for $r_m = 0.94$. Theoretical values of MEVC in Fig.8 are much higher than the measured values in Fig.7. In the present case for fibers of Sample-A with $r_p = 0.5$, the estimated MEVC ~ 1.3 V/cm Oe and is at least three orders of magnitude higher than the measured value in Fig.7(a). In the present case for fibers of Sample-B with $r_m = 0.3$ one expects MEVC of 0.2 V/cm Oe and is two orders of magnitude higher than the measured value in Fig.7(b).

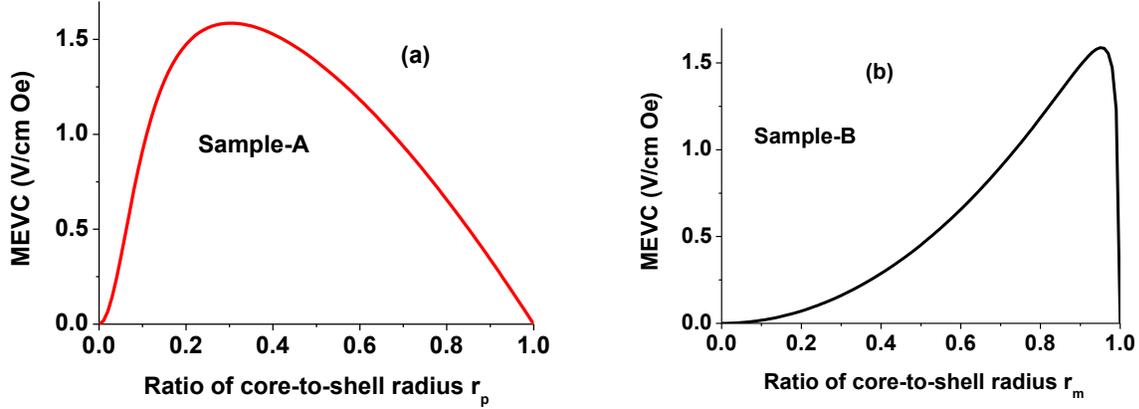

Fig. 8. ME voltage coefficient as a function of ratio of core-to-shell radius for fibers of (a) BTO core and NFO shell and (b) NFO core and BTO shell.

It is clear from the data in Fig.7 and estimates Fig.8 that one has to modify the theory to consider the influence of two important effects on MEVC: (i) magnetic and electric dipole-diploe interactions and (ii) discontinuity in the fibers. For an assembly of nanofibers, we consider the dipole-dipole coupling between magnetic cores or shells and the resulting change in the sample magnetization and in MEVC. For simplicity, we restrict ourselves to the nearest neighbour approximation and assume the geometry of fibers to be uniform and parallel to each other. The angle-dependent free energy includes Zeeman energy and dipole-dipole coupling energy. Assuming the aspect ratio (length-to-radius ratio) of a fiber to be high compared to unity enables one to use the simple approximate expression for interaction energy between two neighbouring magnetic cores,

$$F_{12} = \frac{\mu_0}{2\pi} \frac{VM_{1z}VM_{2z}}{L^2 D} \left[ 1 - \left( 1 + \frac{L^2}{D^2} \right)^{-0.5} \right] \tag{3}$$

with $V$ and $D$ being the magnetic core or shell volume and separation between fibers, respectively. Assuming periodic boundary conditions for the array, one expects the equilibrium angles $\varphi_m$ between the applied bias field and magnetization for neighbor magnetic cores are equal in value and opposite in sign. Minimization of free energy density in $\varphi_m$ results in equilibrium magnetization direction. The piezomagnetic coefficient $q_{33}$ of ferrite core is proportional to Z-component (direction *3*)of magnetization and thus proportional to *cos* $\varphi_m$. Induced voltage due to ME coupling, therefore, decreases because of magnetic dipole interaction. Similarly, equilibrium orientation of PZT polarization will be at an angle $\varphi_e$ to Z-axis (or direction *3*). As a result, piezoelectric coefficient $d_{33}$ decreases due to electric dipole interaction. Thus the ME voltage coefficient in an assembly will be weaker by a factor *t* compared to free standing nanofiber with *t* = *cos* $\varphi_m$ *cos* $\varphi_e$.

The influence of fiber discontinuity on MEVC is considered next. This can be taken into account by assuming an air inclusion along the fiber axis. A fiber with length $L_f$ is assumed to have air inclusions of length $L_a$. Internal magnetic field in a fiber can be expressed in terms of applied field as $\dfrac{H_i}{H} = \dfrac{1}{\dfrac{L_f}{L} + \dfrac{L_a}{L}\dfrac{\mu_f}{\mu_0}}$, where $L = L_f + L_a$, and $\mu_f$ and $\mu_0$ are the permeability of the ferrite and air, respectively. A decrease in the internal magnetic field will result in a decrease in ME coupling strength. Similarly, the relation between the internal electric field and average field across the sample is as follows: $\dfrac{E}{E_i} = \dfrac{1}{\dfrac{L_f}{L} + \dfrac{L_a}{L}\dfrac{\varepsilon_f}{\varepsilon_0}}$. Figure 9 shows the effective fractional MEVC vs. $L_a/L$ for fibers of NFO core-BTO shell. From these results one anticipates a significant reduction in MVC due to fiber discontinuity. An air gap of 17%, for example, is predicted to decrease the MEVC by 3 orders of magnitude. Thus the weakening of the ME coupling strength observed in the MEVC vs H data in Fig.7 for assembled films of fibers could be attributed to air gap (or porosity) in the film.

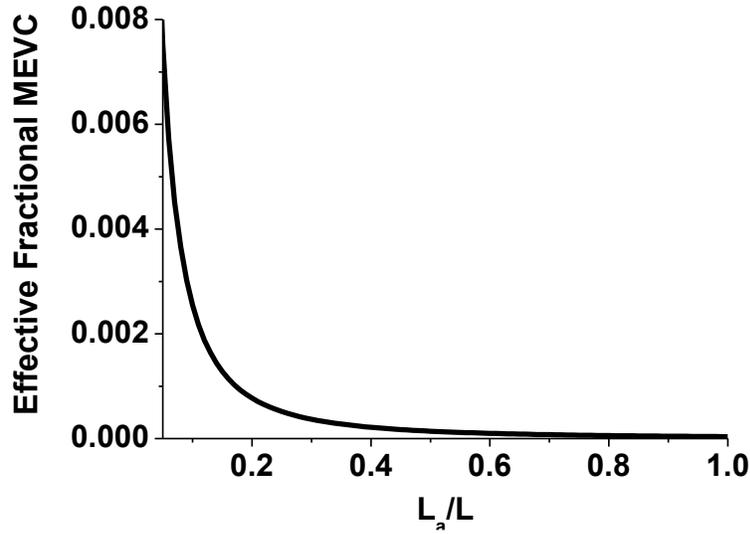

Fig. 9. Effective fractional MEVC as a function of $L_a/L$ for an assembled film of fibers of Sample-A. $L_a$ is the air gap in a fiber of length $L$.

Finally we compare the ME coupling in discs of core-shell fibers and field assembled films with results for similar core-shell nanowires, core-shell particles and also in single phase multiferroics. The data in Fig.4 for magnetic field induced polarization in disks of fibers clearly provide evidence strong ME interactions with 3-6% change in the remnant polarization for H ~ 4 kOe and is comparable to 14-20% reported for discs of core-shell particles of nickel ferrite and BTO [27]. In core-shell fibers of NFO and PZT, ME coupling strength was measured by *H*-induced polarization and found to be 3-4% [34]. Thus the ME coupling measured by *H*-induced polarization in the present system agrees with reported values for NFO-PZT fibers.

The magneto-dielectric effect (MDE) data in Fig.5 indicate strong ME interactions in disks of magnetic-field oriented fibers. The fractional change of 1 to 2% in the permittivity in H=4 kOe in the fibers needs to be compared with MDE in single phase multiferroics and multiferroic nanocomposites. The MDE in single phase multiferroics are generally associated with H induced magnetic phase transitions that result in an induced polarization and a change in the permittivity and is on the order of 1% or less in fields of several Tesla [44]. Thus the MDE in the present system is much stronger than in single phase multiferroics. Past studies on MDE in core-shell nanoparticles of BTO and $ZnFe_2O_4$ reported 1.3% change in $\varepsilon$ for H = 1 T [45]. Similar studies were reported on 50-100 nm diameter nanopyramids of PZT dispersed in a cobalt ferrite matrix. Magneto-capacitance in the system showed a 1.46% increase in H [46]. A composite fiber of nickel zinc ferrite and barium strontium titanate was reported to show magneto-dielectric effect of 14-18% at 1-5 kHz for H = 6 kOe [47].

The MEVC in the present system in Fig.7 is smaller than for thick films of core-shell particles [24-27]. Past efforts on ME characterization of ferrite-ferroelectric core-shell fibers involved the use of magnetic force or piezo force microscopy (PFM) of the nanostructures [48]. An equivalent ME coefficient of 29.5 V/cm Oe was estimated for cobalt ferrite-PZT nanofibers from PFM studies on a single isolated [48]. The primary reason for the low MEVC values is the porosity and defects in the planar films that are seen in the SEM micrographs of Fig.6. The weak ME coupling in the films is in qualitative agreement with the reduction expected due to dipole-dipole interactions and air gap and voids in the assembly.

## 4 . Conclusion

The synthesis of ferrite-ferroelectric core-shell nano-fibers, magnetic field assisted assembly into superstructures and investigations on the magneto-electric coupling are reported. Coaxial nickel ferrite-barium titanate core-shell nanofibers were synthesized by electrospinning. The core-shell structure in the annealed fibers was confirmed with scanning electron microscopy and scanning probe microscopy techniques. The ferromagnetic and ferroelectric nature of the fibers were evident from ferromagnetic resonance and polarization versus electric field measurements.

The ME interactions in annealed fibers were investigated by *H*-induced polarization in discs made from the fibers. The samples showed 3 to 7% change in the remnant polarization under a static magnetic field. Strong ME interactions were also evident from measurements of H-induced variation in dielectric permittivity at 16-22 GHz. The fibers were assembled to form superstructures of *2D* films in a uniform magnetic field or a field gradient. Magnetic field assembled films showed a low-frequency ME voltage coefficient of 450 µV/cm Oe. A model is discussed for low-frequency ME effects that takes into account dipole-dipole interactions and fiber discontinuity in the assembled film. Superstructures of such multiferroic fibers are of interest for use as magnetic sensors and in high frequency devices.

**Acknowledgments**

The research at Oakland University was supported by grants from the Army Research Office (grant no: W911NF1210545) and the National Science Foundation (grant no. ECCS-1337716, 1307714). J. Zhang acknowledges support by the National Natural Science Foundation of China (Grant No. 61503344) and V. M. Petrov acknowledges support by the Russian Science Foundation (project no. 16-12-10158).